# Transverse Thermal Conductivity of REBCO Coated Conductors

M. Bonura, C. Senatore

*Abstract*—REBCO coated conductors (CCs) have the potential to widen considerably the application areas of superconductivity. Quench protection of REBCO-based devices represents one of the major obstacles to this expansion. Thermal conductivity data are mandatory ingredients for quench simulation studies. In particular, the transverse thermal conductivity ($\kappa_T$) of the CC plays a key role in case of quenches in superconducting magnets. In the present work, we report $\kappa_T$ of thermally stabilized REBCO CCs produced by different manufacturers. We have measured $\kappa_T$ on single CCs rather than on stacks of soldered tapes, this excludes spurious contributions in the measurements. We have found that the presence of the stabilizer at the tape edges in Cu-electroplated CCs increases the overall $\kappa_T$. In the absence of this effect, the overall thermal resistance is dominated by the substrate.

## I. Introduction

HIGH-TEMPERATURE SUPERCONDUCTORS (HTS) have been the subject of extensive researches in recent years. The acquired expertise associated to major advances in the fabrication techniques has made REBCO coated conductors (CCs) promising candidates for many applications, such as solenoid and accelerator magnets, fault current limiters, and current cables [1-4].

Thermal properties of REBCO CCs play an important role in the design of HTS-based devices, since they determine the thermal stability. In particular, the thermal conductivity ($\kappa$) of the CC is a mandatory parameter for modeling heat transfer [5,6]. In superconducting windings, both the longitudinal ($\kappa_L$) and the transverse ($\kappa_T$) thermal conductivities, which respectively describe thermal conduction along and across the tape, are needed [5,7]. $\kappa_L$ and $\kappa_T$ assume very different values in REBCO CCs, being $\kappa_L \gg \kappa_T$ [8]. In thermally stabilized CCs, longitudinal heat transfer mainly occurs through the stabilizer (usually Cu). Since thermal conduction in a pure metal is ruled by the electronic contribution, it follows that $\kappa_L$ depends on the purity of the stabilizer as well as on the magnetic field [9,10]. Transverse heat exchange is also influenced by non-Cu layers and in particular by the substrate, which considerably lowers the overall thermal conductivity [11].

The normal zone propagation velocity (NZPV) is a significant parameter for the protection of superconducting devices. Low NZPV values represent one of the main obstacles to the use of REBCO CCs [12,13]. The detection of a quench and the consequent activation of the protection system depends on NZPV. Furthermore, high NZPV values allow limiting the maximum temperature attained during a quench. In case of adiabatic superconducting magnets, the transverse component of NZPV, $v_T$, plays a key role in the quench process [14]. The experimental determination of $v_T$ is much more complex than that of the longitudinal NZPV, $v_L$. Since $v_L$ and $v_T$ are correlated parameters, the latter can be evaluated from $v_L$ and the ratio of the transverse and longitudinal thermal conductivities: $v_T = v_L \sqrt{\kappa_T / \kappa_L}$ [5].

Very few experimental works on the thermal conductivity of REBCO tapes have been reported in the literature [4,8]. Furthermore, transverse thermal properties have never been examined on single CCs but only on stacks of tapes [8,15]. In a recent work, we have investigated the longitudinal thermal conductivity of industrial REBCO CCs [9]. In this manuscript, we report the transverse thermal conductivity of REBCO tapes produced by different manufacturers. These data can be of particular interest for the magnet design community.

## II. Samples

We have investigated REBCO CCs from five different manufacturers, namely American Superconductor (AMSC), Bruker HTS (BHTS), Fujikura, SuNAM, and SuperPower. The tape layout as well as the constituent materials may differ from one manufacturer to the others. Table I summarizes some of the CCs' properties interesting for this study.

AMSC and BHTS use Ni5at.%W (NiW) and stainless steel for the substrate, respectively. The other manufacturers employ Hastelloy. Each manufacturer has his own recipe for the buffer layers (details in Table I). The thickness of the REBCO layers is about 1-2 μm for all the CCs [2]. A silver layer of about 1-2 μm is deposited over the REBCO layer for protection against the moisture from the environment. In the tape manufactured by SuperPower, a second Ag layer is also present below the substrate. All the CCs investigated in this work have been stabilized with Cu either by electroplating (BHTS, SuNAM, and Superpower) or by lamination (AMSC and Fujikura).

The authors acknowledge the financial support from the Swiss National Science Foundation (Grant N. PP00P2-14467 and 51NF40-144613). Research also supported by FP7 EuCARD-2 http://eucard2.web.cern.ch. EuCARD-2 is co-funded by the partners and the European Commission under Capacities 7th Framework Programme, Grant Agreement 312453.

The authors are with the Department of Applied Physics (GAP) and the Department of Condensed Matter Physics (DPMC), University of Geneva, Quai Ernest Ansermet 24, CH-1211 Geneva, Switzerland (e-mail: marco.bonura@unige.ch; carmine.senatore@unige.ch).

TABLE I
DETAILS ON THE LAYOUT OF INVESTIGATED COATED CONDUCTORS

| Manufacturer | Substrate | Buffer Layers | Stabilization |
|---|---|---|---|
| AMSC | Ni5at.%W | $Y_2O_3$/YSZ/$CeO_2$ | Cu laminated |
| BHTS | Stainless Steel | YSZ/$CeO_2$ | Cu electroplated |
| FUJIKURA | Hastelloy | $Al_2O_3$/$Y_2O_3$/ MgO/$CeO_2$ | Cu laminated |
| SUNAM | Hastelloy | $Al_2O_3$/$Y_2O_3$/ MgO/$LaMnO_3$ | Cu electroplated |
| SUPERPOWER | Hastelloy | $Al_2O_3$/$Y_2O_3$/ MgO/$LaMnO_3$ | Cu electroplated |

The CC from Fujikura has Cu just on one face of the tape. AMSC uses two Cu strips, one for each tape face, soldered together at the CC edges. Electroplated CCs have Cu all around the tape, edges included.

In Table II, we have reported other CCs' properties of interest for this study, namely the overall thickness and the thickness of substrate and stabilizer, and the Cu residual resistivity ratio $RRR \equiv R(273\ K)/R(4.2\ K)$ obtained by measuring the electrical resistivity on Cu samples extracted from the CCs [9]. We have determined the layers' dimensions by optical microscopy.

TABLE II
PROPERTIES OF INVESTIGATED COATED CONDUCTORS

| Manufacturer | Overall Thickness | Substrate Thickness | Copper Thickness | Copper $RRR$ |
|---|---|---|---|---|
| AMSC | 200 μm | 75 μm | 100 μm | 19 |
| BHTS | 150 μm | 100 μm | 30 μm | 17 |
| FUJIKURA | 160 μm | 75 μm | 75 μm | 59 |
| SUNAM | 110 μm | 60 μm | 36 μm | 61 |
| SUPERPOWER | 100 μm | 50 μm | 40 μm | 42 |

## III. EXPERIMENT

### A. Experimental Method

We have investigated the thermal conductivity using a laboratory-made setup specifically designed for measurements on superconducting wires and tapes. $\kappa_T$ is defined as

$$\kappa_T = \frac{Q}{\Delta T}\frac{t}{A}, \qquad (1)$$

where $Q$ is the heat supplied to one face of the tape, $\Delta T$ the temperature gradient between the two faces, $t$ the thickness of the CC and $A$ the surface of the sample. For our setup, $A$ is typically $\approx 20$ mm$^2$.

Transverse thermal conductivity measurements in single CCs require a high sensitivity of the experimental apparatus. The reduced thickness of the tape can make $\Delta T$ so small that its experimental determination can be challenging [8]. We have taken specific precautions for the samples' preparation. In Fig. 1, we report a schematic drawing of the sample holder. The CC is sandwiched between two copper leads glued on the sample's faces by GE varnish. This ensures good thermal contact over the entire tape surface. The leads are in high-$RRR$ Cu and the temperature gradient along the leads is negligible with respect to $\Delta T$. A hole is present on each Cu lead in correspondence of the center of the sample. This allows gluing a Cernox bare chip, used as thermometer, directly on the tape face. Two calibrated Cernox thermometers are also present in the setup, one on the heat sink and one on the heater holder. During the measurement, the heat-sink temperature is kept constant. When current is supplied to the heater, energy is generated by Joule effect and a temperature gradient is established between the two sample faces. $\Delta T$ is measured once the steady-state heat flow has been reached. Typically $\Delta T$ is of the order of 10 mK.

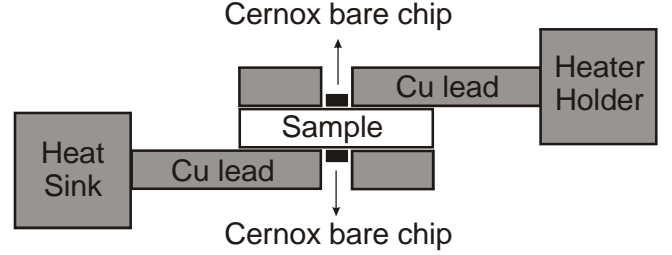

Fig. 1. Sketch of the sample holder.

The high values for the derivative of the resistance vs temperature curve of the Cernox bare chips ($dR/dT > 1$ kΩ/K for $T$<10 K), permit an easy detection of temperature variations as small as 0.1 mK. Bare chips are mounted on the sample and their calibration is performed when the sample is at the thermal equilibrium with the bath. This removes possible errors associated to the thermal contact between the chips and the sample surface due to the GE varnish. The value of $dR/dT$ becomes smaller as temperature increases, and this leads to a growing uncertainty in the measurement. For this reason, we have investigated $\kappa_T$ up to $T \approx 25$-30 K. In the case of the CCs from BHTS and SuperPower, smaller $\kappa_T$ values (larger $\Delta T$) have allowed us to extend the investigation up to $T \approx 35$ K.

### B. Measurement Error

The measurement uncertainty is mostly due to the evaluation of the ratio $t/A$, the associated maximum error being about 30%. The error on $Q$ is smaller than the 0.05%, since we have taken specific precautions for limiting convective, conductive and radiating heat losses [16]. Errors associated to $\Delta T$ increase with the temperature. This is a consequence of the characteristic $R(T)$ curve of Cernox bare chips. The maximum error associated to $\Delta T$ in the range of temperatures investigated is 15%. For $T$<10 K this percentage is smaller than 5%.

## IV. RESULTS AND DISCUSSION

In Fig. 2, we report the temperature dependence of the transverse thermal conductivity for all the investigated samples. CCs from different manufacturers show different $\kappa_T$ values because of the different layouts and materials used. For all the tapes, we have observed an increase of $\kappa_T$ with the temperature.

Transverse thermal conduction in layered CCs can be analyzed with a formalism analogous to the case of electrical resistances connected in series. In Fig. 3, we show a simplified illustration of two different categories of CCs with the associated thermal/electrical circuit. $R_{O-L}$ indicates the thermal resistance associated to all the materials other than Cu and substrate. The configuration shown in Fig. 3 (a) is particularly appropriate to the CC from Fujikura, where Cu has been laminated only on one side of the tape. The overall thermal resistance is the sum of the resistances associated to each layer:

$$R_{tot} = \sum_i \frac{1}{\kappa_i} \frac{t_i}{A} + R_{Int}, \quad (2)$$

where $\kappa_i$ is the thermal conductivity associated to the $i^{th}$ layer, whose thickness is $t_i$. In (2) we have also included the term $R_{Int}$ associated to the interface thermal resistance due to a non-perfect connection between different layers.

In Fig. 3 (b) we show the typical cross-section of a Cu-electroplated CC. The stabilizer is deposited all around the tape. Thus, the contribution to the overall thermal transport from the Cu at the tape edges has to be considered, too. Transverse thermal transport in the Cu-laminated CC from AMSC can also be schematized following the circuit in Fig. 3 (b). In this CC, the upper and lower Cu strips are jointed together at the tape edges by a metal solder.

In the case of the CC from Fujikura, the absence of Cu on the tape edges allows us to use (2) for analyzing the experimental results of Fig. 2. $\kappa$ values of the Cu present in the CC can be deduced from longitudinal thermal conductivity measurements, since $\kappa_L \approx \kappa_{Cu} s_{Cu}$, where $s_{Cu}$ is the cross-section fraction occupied by the stabilizer [9]. The temperature dependence of $\kappa_L$ for the CCs investigated in this work have been reported in Ref. [9]. It is worth analyzing the thermal behavior of the CC at temperatures close to values corresponding to the current sharing between superconductor and metal layers. In the case of REBCO CCs operating at $T= 4.2$ K, this typically occurs between 25K and 35 K [9,17]. We deduce from Ref. [9] that the value of $\kappa_{Cu}$ at $T= 25$ K is $\approx 1740$ WK$^{-1}$m$^{-1}$. Lu et al. have reported on thermal conductivity of Hastelloy in a wide range of temperatures [18]. At $T= 25$ K they have found $\kappa_{Hastelloy} \approx 5$ WK$^{-1}$m$^{-1}$. In the CC from Fujikura, Cu and Hastelloy layers have the same thickness ($t=75$ μm). From (2), it follows that $R_{Hastelloy}/R_{Cu}= \kappa_{Cu}/\kappa_{Hastelloy}$ and we can straightforwardly deduce that the contribution from the Cu layer to the overall thermal resistance can be neglected, being more than 2 orders of magnitude smaller than the one associated to the substrate. Using $\kappa(T)$ values of Hastelloy from Ref. [18] and expression reported in (2), we can calculate the expected value of $\kappa_T(T)$ for the Fujikura tape, in the hypothesis $R_{Int}= R_{O-L}=0$. The calculated curve is reported in Fig. 2 as a continuous line. We have found a very good agreement between experimental and calculated data in the entire range of temperatures investigated. This indicates that for the CC from Fujikura, transverse thermal resistivity is dominated by the substrate contribution.

The same analysis cannot be performed for the CCs from the other manufacturers. This is a consequence of the contribution to the transverse thermal transport from the Cu (solder in the case of AMSC) on the tape edges. An illustrative example comes from the analysis of the results for the SuNAM tape. We can schematize thermal transport as in the circuit of Fig. 3 (b). Considering $\kappa$ values for the Hastelloy reported in Ref. [18] and using the geometrical specifications of the investigated sample ($t_{Hastelloy} \approx 60$ μm, $A \approx 20$ mm$^2$), we find: $R_{sub} \approx 0.6$ W$^{-1}$K at $T=25$ K. We can also estimate the thermal resistance associated to the Cu present on the edges of the tape, $R'_{Cu}$. From the results reported in Ref. [9] for the CC from SuNAM, we deduce that $\kappa_{Cu} \approx 1910$ WK$^{-1}$m$^{-1}$ at $T=25$ K. We have measured by optical microscopy the average thickness of the lateral Cu and it resulted to be $\approx 40$ μm (2 x 20 μm). The length of the sample is $\approx 5$ mm, and thus we infer that $R'_{Cu} \approx 0.2$ W$^{-1}$K. This would indicate that at $T= 25$ K about 2/3 of the overall thermal current passes through the Cu present at the edges of the CC. It is noteworthy that the thermal conductivity of materials composing the CC, including Cu and Hastelloy, are temperature dependent [9,10,11,18]. This could change the fractional amount of heat

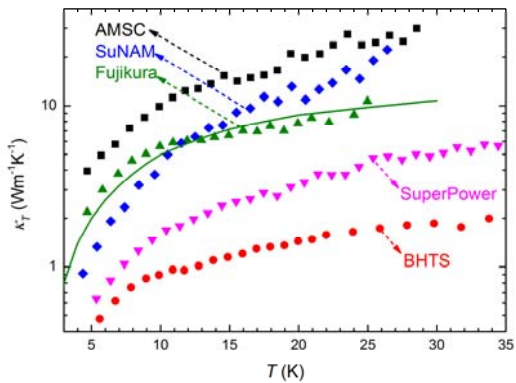

Fig. 2. Temperature dependence of the transverse thermal conductivity of REBCO coated conductors from different manufacturers. The continuous line indicates the expected curve for the tape from Fujikura.

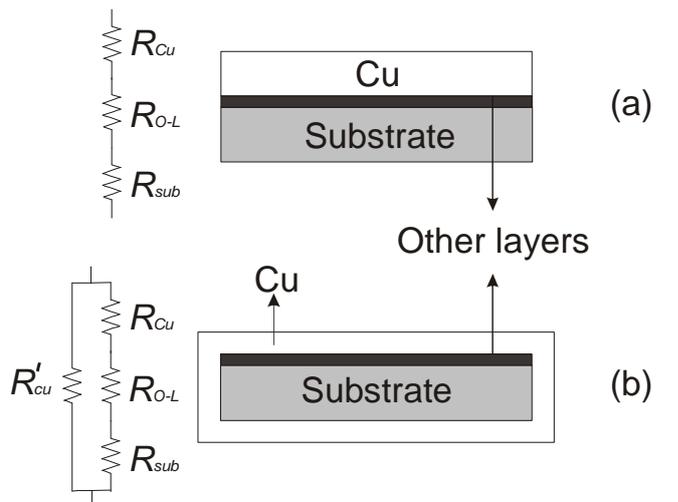

Fig. 3. Schematic drawing of stabilized CCs with associated equivalent thermal circuits. In (a), Cu is present only on one face of the tape and no parallel between thermal resistances is present. This is for example the case of the CC from Fujikura. In (b), Cu is present all around the tape, creating a parallel of thermal resistances. Case (b) particularly applies to Cu-electroplated tapes.

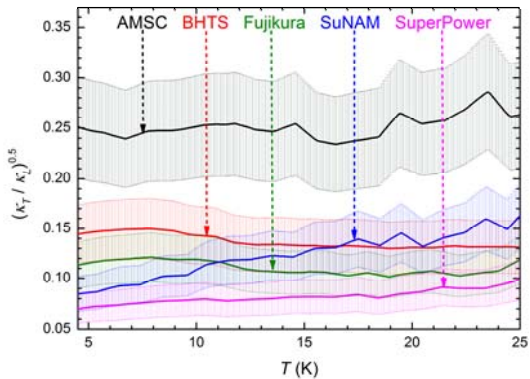

Fig. 4. Temperature dependence of the square root of the ratio $\kappa_T/\kappa_L$. $\kappa_L$ values of the investigated CCs have been extracted from [9]. Error bars are mostly determined by the uncertainty in $\kappa_T$ measurements.

passing through the lateral copper on varying the temperature.

We have shown here that the contribution from the lateral Cu to the overall $\kappa_T$ cannot be generally neglected in the case of electroplated Cu. Nevertheless, a quantitative analysis of $\kappa_T$ is hindered by different factors. Fluctuations of the Cu thickness at the sample edges impede a precise determination of the dimensions of the lateral Cu strips. Furthermore, variation in morphology of Cu grains may arise from the deposition process and this has an influence on the local electron transport properties [19,20]. This confirms the importance of direct measurements in the determination of thermal properties of CCs.

We could not perform a quantitative analysis of $\kappa_T(T)$ for the Cu-laminated CC from AMSC due to the lack of information on the material used to solder the two Cu strips on the edges.

From Fig. 2, we notice that the CC from AMSC has the highest $\kappa_T$ values. In this CC, the substrate is made by NiW. The temperature dependence of the thermal conductivity of NiW has been reported in Ref. [21]. The better transverse thermal conduction properties of the CC from AMSC are correlated to the large amount of Cu used by this manufacturer (see Table II). In fact, the high thermal conductivity of Cu increases the overall $\kappa_T$.

The tape from BHTS exhibits the lowest $\kappa_T$. This is a consequence of Cu thickness and of the use of stainless steel (SS) as substrate material, being $\kappa_{SS} < \kappa_{Hastelloy}$ [11].

We have demonstrated that the contribution to thermal conduction of the Cu on the edges of the tape may affect the overall transverse thermal conduction. For the sake of completeness, we would like to point out that the field-induced reduction of $\kappa$ of Cu should also be taken into account, in case of use of the HTS in magnetic fields. Field-induced effects could considerably increase $R'_{Cu}$, especially in case of high-$RRR$ Cu [9,10].

In Fig. 4 we plot the square root of the ratio $\kappa_T/\kappa_L$, as calculated from the results of this work and $\kappa_L$ data reported in [9]. Without considering the contribution of the impregnation to the overall thermal transport in a magnet, these values can give valuable information on the anisotropy of the NZPV. Within the experimental uncertainty, most of the CCs have comparable values for $(\kappa_T/\kappa_L)^{0.5}$. Only the CC from AMSC shows larger values with respect to the other CCs. This result is related to the high $\kappa_T$ measured in this CC.

## V. CONCLUSION

We have studied the transverse thermal conduction in REBCO coated conductors from different manufacturers. The use of an experimental setup specifically designed for thermal studies on technical superconductors has allowed us to measure $\kappa_T$ on single CCs and not on stacks of soldered tapes. This eliminates possible errors in the evaluation of $\kappa_T$ due to the influence of the solder. We have shown that the main contribution to the transverse thermal resistance comes from the substrate in the case of the CC from Fujikura, where Cu is present only on one side of the tape. For the CCs produced by the other manufacturers, Cu (solder for AMSC) on the tape edges opens parallel channels to thermal conduction that lower the overall thermal resistance. The comparison of transverse and longitudinal thermal conductivities of the CCs has allowed us to shed light on the anisotropy of the NZPV.

ACKNOWLEDGMENT

We acknowledge Christian Barth and Ciro Calzolaio for useful discussions, Damien Zurmuehle and Eugenie Gallay for technical support.